\documentclass{sig-alternate}
\usepackage{subfigure}
\usepackage{xspace}
\usepackage{hyperref}
\usepackage{xcolor}
\hypersetup{
    colorlinks,
    linkcolor={red!50!black},
    citecolor={blue!50!black},
    urlcolor={blue!80!black}
}

\newcommand{\eg}{\textit{e.g.}\xspace}

\newcommand{\connectedness}{\textit{largest cluster}\xspace}
\newcommand{\simple}{\textit{simple}\xspace}
\newcommand{\heft}{HEFT\xspace}

\begin{document}

\title{On transaction parallelizability in Ethereum}

\numberofauthors{1}
\author{
\alignauthor
Nadi Sarrar
\vspace{1.5mm}
\email{nadi@fibium.com}
}

\maketitle

\begin{abstract}
Ethereum clients execute transactions in a sequential order prescribed by the consensus protocol. This avoids conflicts due to different transactions reading or modifying the state of the same accounts. Sequential execution constitutes a safe and conservative approach to blockchain transaction processing which forgoes running transactions in parallel even when doing so would be beneficial and safe, \eg, when there is no intersection in the sets of accounts that the transactions touch. In particular, the possibility of unpredictable conflicts greatly complicates the implementation of parallel transactions. This calls for a study quantifying the occurrence of such conflicts and analyzing the speedup potential from parallelization under such conditions.

In this work we take a step in this direction and conduct an empirical study based on all blocks and transactions in Ethereum to date to develop an understanding of actual transaction execution constraints. Our results suggest that the opportunity of safely running transactions in parallel is improving over time. Notably, a simple scheduler already achieves average speedups of 1.77 and 1.94 in years 2017 and 2018, respectively. Using the more advanced Heterogeneous Earliest Finish Time (\heft) scheduler we observe even higher gains.

We believe that there are practical ways to leverage these insights in production, including \textit{(i)}~by easy parallelizability~\cite{eip:648} and access lists~\cite{spec:accesslist}, and \textit{(ii)}~by speculatively running transactions in parallel with a mechanism to resolve conflicts as they occur~\cite{DBLP:journals/corr/DickersonGHK17,DBLP:journals/corr/abs-1809-01326}. The results of this work are more closely related to~\textit{(i)} as we assume future knowledge of each transaction's access list (and processing time, in case of \heft) and use it to determine a schedule that never aborts.
\end{abstract}

\section{Methodology}
We use a modified version of go-ethereum~v1.8.19\footnote{\url{https://github.com/nadisarrar/go-ethereum/}} which maintains for each transaction a record of all accounts which are relevant during its execution, including the \texttt{From} account, the \texttt{To} account (except for \texttt{create} transactions), all accounts created directly or indirectly, and all accounts that the transaction interacts with, including value transfers, method invocations, and any access to state such as \texttt{balance} and \texttt{codeHash}.

We feed these transaction records into three different simulators: one that is based on transaction graph properties, which we call \connectedness, and two transaction schedulers, \simple and \heft. We rely on \texttt{gasUsed} as an estimator of transaction processing times -- this is not always accurate but produces repeatable results which are independent of specific client implementations, hardware capabilities, network performance, etc.

\subsection{Largest Cluster}
Transactions are represented as vertices in a graph, with edges between a pair if and only if they have at least one account in their access lists in common. The \connectedness metric is based on the largest (in total \texttt{gasUsed}) disjoint set of transactions in that graph. We calculate the total processing time as the time it takes to sequentially process that largest disjoint subset\footnote{There may be transactions belonging to the same transitively connected group that can safely be processed in parallel. The \connectedness simulation therefore leaves room for optimization.}. We assume that there are infinite threads available so that we can allocate a dedicated thread to each remaining subset and therefore make sure that the smaller sets finish no later than the largest one. This metric serves as a point of comparison for the other metrics that are limited in the number of available threads.

\subsection{Simple Scheduler}
The \simple scheduler processes transactions in batches of size equal to the number of threads. Batches are allowed to contain only non-conflicting transactions. This means that batches may contain fewer transactions than there are threads available, and a batch only completes when its longest running transaction completes. For these reasons some threads may be left underutilized. Since the dataset contains access lists for each transaction, the scheduler can determine a schedule that is guaranteed to not conflict.

\subsection{HEFT Scheduler}
\label{sec:heft}
\heft~\cite{heft} is a heuristics-based task scheduling algorithm that minimizes overall completion time. The implementation of \heft used in this study is available at~\cite{git:heft}. We configure \heft to use the consensus ordering of transactions as well as their access lists as precedence constraints, and the \texttt{gasUsed} metric as an approximation of processing time. Unfortunately, the python implementation of \heft used in our experiments failed for some large blocks. As a work around, we compute schedules for at most 32 transactions at a time, which means that larger blocks require multiple runs of \heft. We impose the same 32 transaction limitation in the \simple and \connectedness experiments as well to ensure a fair comparison.

\begin{figure*}[!ht]
    \centering
    \subfigure[Comparison of \connectedness and \simple.]{
        \includegraphics[width=.485\textwidth]{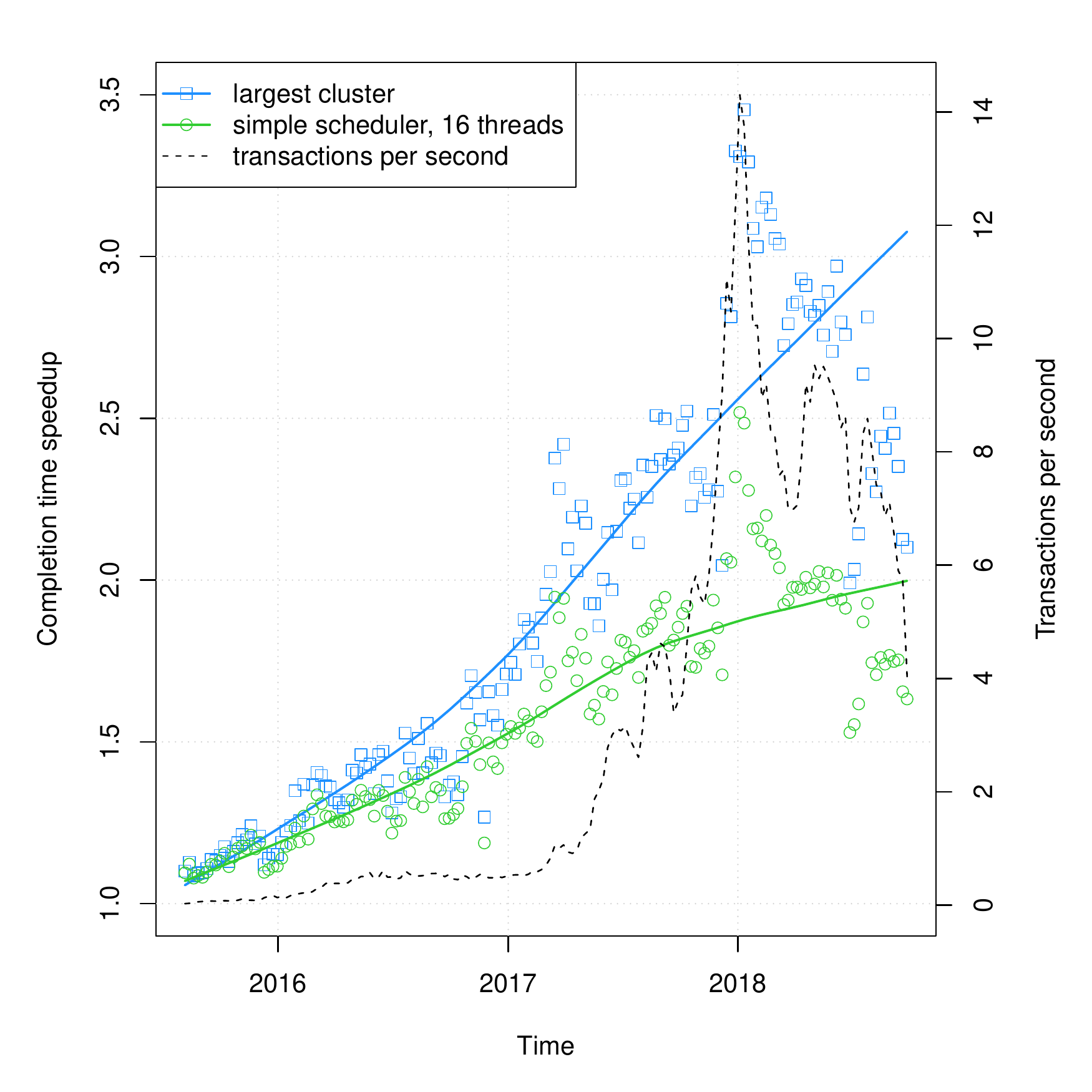}
        \label{fig:unbatched}
    }
    \subfigure[Comparison with a limit of 32 transactions per schedule.]{
        \includegraphics[width=.485\textwidth]{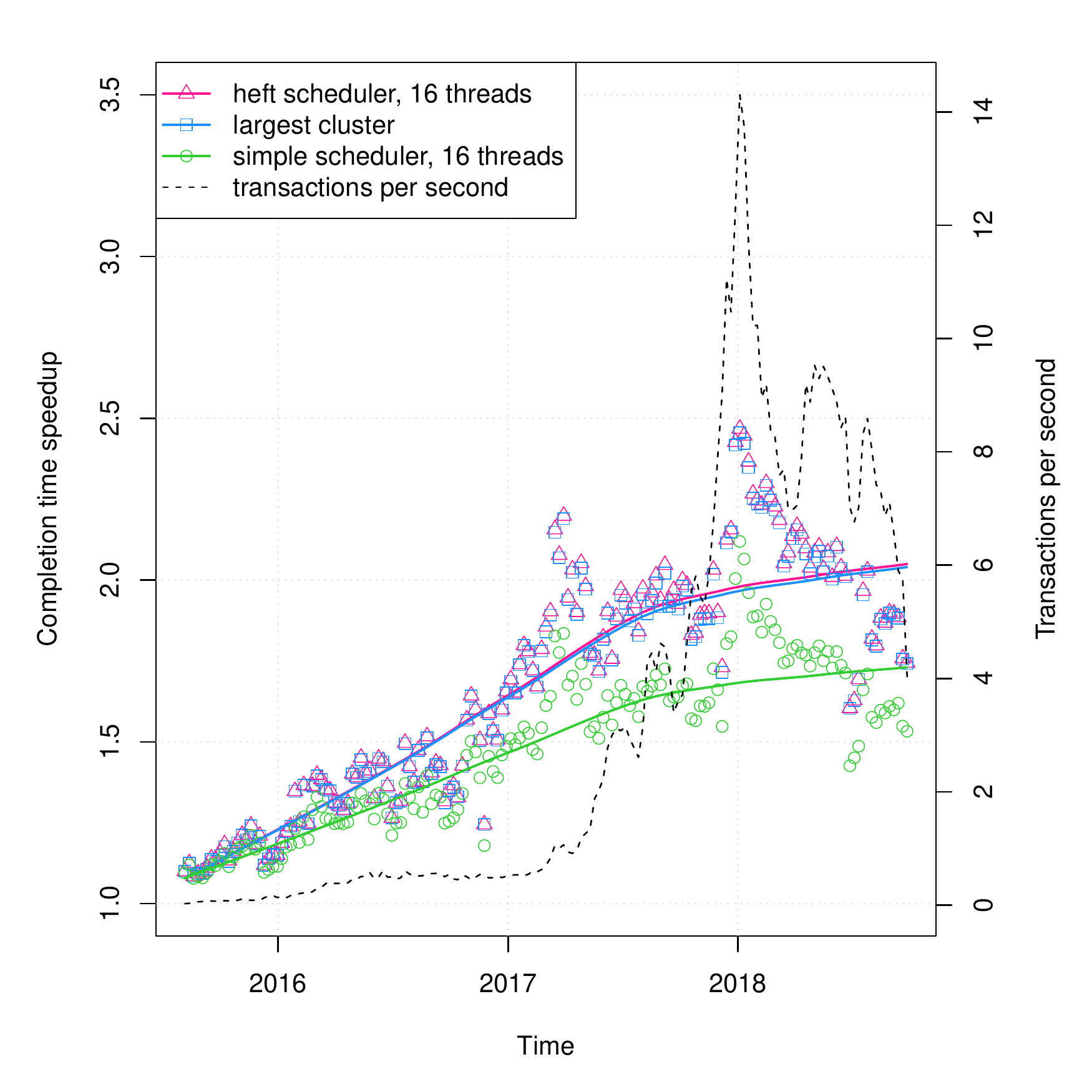}
        \label{fig:batched}
    }
    \caption{Parallelization gain as achieved by \connectedness, \simple, and \heft.}
    \vspace{-0.40em}
    \label{fig:results}
\end{figure*}

\section{Results}
\sloppy{In Figure~\ref{fig:unbatched} we compare the performance of the \connectedness approach to that of the \simple scheduler approach. There are three main take-aways. The direction of the LOESS regressions (solid lines) suggests that the potential for parallelizability improves over time. However, we observe a declining performance in 2018 which appears to coincide with a decreasing transaction rate. We speculate that the more transactions there are, the better they can be parallelized, due partly to there being little room for parallelization if a block contains few transactions, and partly to an increase in the number of popular accounts (smart contracts) and disjoint user groups generating distinct clusters of transactions. Lastly, while the \simple scheduler performs worse than \connectedness, it still shows promising gains overall.}

Figure~\ref{fig:batched} compares the \heft scheduler to the other two simulations. These experiments use a cap of 32 transactions per schedule because of a limitation in the \heft implementation~(Sec.~\ref{sec:heft}). This results in decreased performance of \connectedness as well as \simple. The overall trends however remain largely the same. \heft outperforms \simple, reaching the same levels as \connectedness.

In summary, while average thread utilization remains modest compared to the number of available threads, all three simulations demonstrate that notable gains are achievable with parallelization.

\section{Future work}
This study assumes that we have future knowledge of the accounts that a transaction touches. While we may eventually be able to use access lists~\cite{spec:accesslist} in an upcoming version of Ethereum for that purpose, further research on scheduling algorithms which relax this assumption will be needed.

In case of \heft, we further assume that \texttt{gasUsed} serves as an approximation of a transaction's processing time. This is not feasible in practice because \texttt{gasUsed} only becomes available \textit{after} a transaction was run, as part of the transaction receipt. Future research may investigate the (un)predictability of a transaction's processing time.

This work makes no claims as to where resource bottlenecks in Ethereum currently lie. We hope our work complements studies investigating constrained resources such as networking and disk I/O, as well as work on alternative concurrent state database data structures.

Finally, we hope this research direction will prove useful in the current phases of Ethereum Serenity development, where accounts are assigned to shards (groups), and cross-shard communication (messages between groups) incurs some overhead. Intuitively, a greater parallelizability may suggest a lesser need for cross-shard communication, depending on our ability to correctly identify the clusters in real time as accounts are created and assigned to shards, and the rate at which those assignments change.

\small
\bibliographystyle{ieeetr}
\bibliography{paralleltx}

\begin{thebibliography}{1}

\bibitem{eip:648}
V.~Buterin, ``Easy parallelizability,'' {\em Ethereum Improvement Proposal
  648}, June 2017.

\bibitem{spec:accesslist}
``Ethereum sharding specification proposal: Access lists.''
  \url{https://github.com/ethereum/sharding/blob/master/docs/doc.md#access-list},
  accessed 01/2019.

\bibitem{DBLP:journals/corr/DickersonGHK17}
T.~D. Dickerson, P.~Gazzillo, M.~Herlihy, and E.~Koskinen, ``Adding concurrency
  to smart contracts,'' {\em CoRR}, vol.~abs/1702.04467, 2017.

\bibitem{DBLP:journals/corr/abs-1809-01326}
P.~S. Anjana, S.~Kumari, S.~Peri, S.~Rathor, and A.~Somani, ``An efficient
  framework for concurrent execution of smart contracts,'' {\em CoRR},
  vol.~abs/1809.01326, 2018.

\bibitem{heft}
H.~Topcuoglu, S.~Hariri, and M.-Y. Wu, ``Performance-effective and
  low-complexity task scheduling for heterogeneous computing,'' {\em IEEE
  Transactions on Parallel and Distributed Systems}, vol.~13, March 2002.

\bibitem{git:heft}
M.~Rocklin, ``{HEFT} implementation in {P}ython.''
  \url{https://github.com/mrocklin/heft}, 2013-2017.

\end{thebibliography}
\end{document}